\pdfoutput=1

\documentclass[11pt,a4paper]{article} 
\usepackage{jcappub} 

\usepackage{amsmath}
\usepackage{amsfonts}
\usepackage{graphicx}
\usepackage{hyperref}
\usepackage{natbib}
\usepackage{bm}

\usepackage{varioref}
\usepackage{amssymb}
\usepackage{comment}

\title{Predictions in multifield models of inflation}
\author{Jonathan Frazer}
\affiliation{Astronomy Centre, University of Sussex, Brighton BN1 9QH,
United Kingdom}

\abstract{This paper presents a method for obtaining an analytic expression for the density function of observables in multifield models of inflation with sum-separable potentials. The most striking result is that the density function in general possesses a sharp peak and the location of this peak is only mildly sensitive to the distribution of initial conditions. A simple argument is given for why this result holds for a more general class of models than just those with sum-separable potentials and why for such models, it is possible to obtain robust predictions for observable quantities. As an example, the joint density function of the spectral index and running in double quadratic inflation is computed. For scales leaving the horizon 55 $e$-folds before the end of inflation, the density function peaks at  $n_{s}=0.967$ and $\alpha=0.0006$ for the spectral index and running respectively.}


\begin{document}

\maketitle
\flushbottom


\section{Introduction}

The observable consequences of the simplest models of inflation with only one canonical field are quite well understood but ideas from fundamental physics seem to motivate models with more than one active field.\footnote{As described in Ref.~\cite{Cicoli:2012cy}, although it remains difficult in string theory to achieve a sufficiently light scalar to give rise to inflation, once this is managed, the mechanism which gives rise to one light scalar, tends to give rise to many.} Models involving more than one light field, often referred to as ``multifield models" allow for much richer inflationary behaviour and consequently making a prediction for observables in models of this kind is more complex. Although the phenomenology of these models is well studied, and the search for observational signatures such as local non-Gaussianity is an active area of research, at present it is not clear what even the simplest multifield models actually predict.

One characteristic which makes computing observables more challenging is the fact that the primordial curvature perturbation evolves on superhorizon scales. In order to understand the possible signatures of multifield inflation it is essential to be able to precisely track the evolution after horizon crossing. Fortunately this is well studied and a number of techniques exist \cite{Starobinsky:1986fxa,Lyth:1984gv,Sasaki:1995aw,
	Salopek:1990jq,Sasaki:1998ug,Wands:2000dp,Lyth:2005fi,Rigopoulos:2004gr,Rigopoulos:2005xx,Yokoyama:2007uu,Yokoyama:2007dw,Yokoyama:2008by,Mulryne:2009kh,Mulryne:2010rp,Amendola:2001ni,GrootNibbelink:2001qt,Lalak:2007vi,Peterson:2010np,
	Peterson:2010mv,Peterson:2011yt,Achucarro:2010da,Avgoustidis:2011em,Lehners:2009ja,Ringeval:2007am,Martin:2006rs,Huston:2009ac,Huston:2011vt}. However, there is another important distinction between single field models of inflation and models with multiple active fields which is the sensitivity to initial conditions. 

Models involving only one field are essentially insensitive to initial conditions; some minimum number of $e$-folds of inflation is required to solve cosmological problems such as the flatness and homogeneity, but provided the total amount of inflation is a few $e$-folds more than this, the observational consequences are independent of how much inflation actually takes place.\footnote{An important exception is the situation where there is more than one metastable vacuum. This situation certainly is sensitive to initial conditions. The discussion in this paper applies when only one minimum is relevant.} The primordial curvature perturbation on a given scale is uniquely determined by the value of the inflaton at the moment that scale left the horizon. If we take the largest scales we observe to have left the horizon say 55 $e$-folds before the end of inflation, our observations are generally insensitive to what happened before then, be it another 20 $e$-folds of inflation or 200.

\begin{figure}[t]
\centering
\includegraphics[width=12cm]{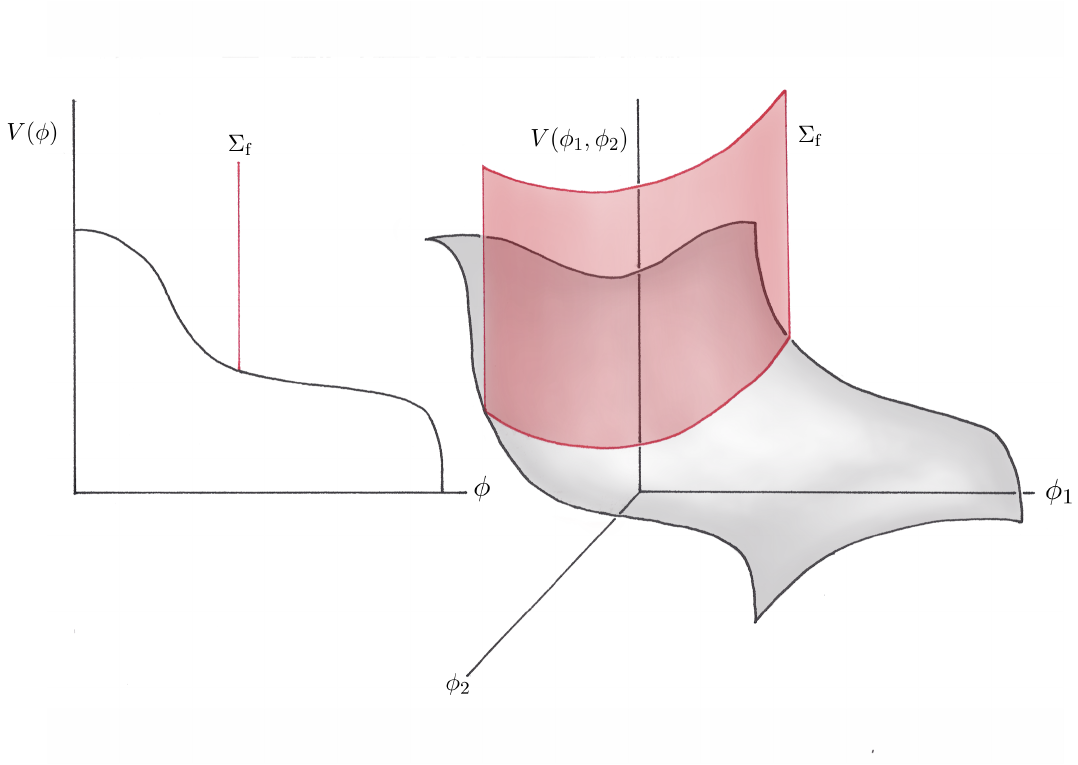}
\caption{Sketch showing the horizon crossing surface in a single field inflationary potential and a two-field model. For a single field model, the horizon crossing surface is a single point, meaning that observables are approximately insensitive to inflationary dynamics prior to crossing this point. For the case of the two-field model, without knowledge of initial conditions one can not identify a single inflationary trajectory that corresponds to what we observe. One must therefore consider all possible inflationary trajectories.}
\label{fig:surface}
\end{figure}

When there is more than one active field the situation changes. As illustrated in Fig.~\ref{fig:surface},  if the model involves $N_{\rm f}$ inflationary fields, instead of there being only one possible value of the inflaton at horizon crossing, now there is an infinite set of possible locations in field space forming an $N_{\rm f}-1$ dimentional hypersurface.\footnote{Discussed in more detail in section \S\ref{sec:sep}, this assumes only one trajectory passes through a given point in field space.} Without specifying initial conditions, it is not possible to say which inflationary trajectory corresponds to our observable universe and since different trajectories will in general give rise to different values for observables, the model is only as predictive as the volume in the space of observables permitted by the model. If we are to seriously confront multifield models of inflation with observation, then it is of paramount importance that this problem is overcome.

The obvious question then is whether a description of initial conditions can be derived within the framework of the model. One can hope that an ultraviolet complete theory of inflation will provide information on the initial state but calculations along this line are clearly well beyond our current understanding of fundamental physics. A more promising approach is to consider how chaotic inflation populates the potential. Although this issue has received some attention in the past for the case of single field inflation (see for instance Refs.~\cite{Lyth:2006gd,Seery:2009hs} and references therein), a general description for the multifield case currently does not exist. 

Ultimately the most problematic aspect of this question is what is often referred to as the measure problem (see Ref.~\cite{Freivogel:2011eg} for a recent review). Often described in the context of tunnelling between metastable vacua but nevertheless still of critical importance even to much simpler models~\cite{Frazer:2011br}, the issue is that one must choose a measure when addressing what proportion of an infinite space-time corresponds to a particular choice of an infinite set of initial conditions. 

Despite the significant challenge of computing initial conditions, the aim of this paper is to show that at least for some models it is still possible to make robust predictions for observables. A general method is not given here. Instead, by considering the subclass of multifield models where there is no couplings between the fields, it is shown that the density function for observables can be computed analytically given an initial density function on field space. It turns out that under mild assumptions, the density function of observables is strongly peaked. This characteristic is largely determined by the geometry of the potential and can be comparatively insensitive to the initial density function on field space. This makes clear how the prediction for observables is (marginally) affected by the choice of measure, leading one to conclude that only certain choices will significantly change the result in comparison to the resolution of current and future observations such as Planck.

The approach taken in this paper utilises the fact that for sum-separable potentials, provided isocurvature modes have decayed before the end of inflation, observables $o$ may be expressed solely in terms of quantities evaluated at horizon crossing. As is often the case, in what follows this will be referred to as the horizon crossing approximation \cite{Kim:2006te}.\footnote{Despite the potentially misleading name, it should be noted that this approximation \emph{does} account for superhorizon evolution, as will be discussed in \S\ref{sec:sep}.} The use of this approximation means that all the information required to compute observables for \emph{all} possible inflationary trajectories is contained in a single hypersurface $\Sigma_{\rm f}$ parameterised by $N_{\rm f}-1$ variables $\{\theta_{1},..,\theta_{N_{{\rm f}-1}}\}$. The idea then is that by specifying the density function $f(\theta_{1},..,\theta_{N_{{\rm f}-1}})$ on the hypersurface, conservation of probability implies it is possible to compute the resulting density function for observables $p(o_{1},...,o_{n})$. For instance, consider the case where the hypersurface is 1-dimensional: Let $\theta$ be a continuous random variable with probability density $f(\theta)$. Defining $o\equiv o(\theta)$, provided the function is bijective, the probability density of $o$ is given by $p(o)$ where
\begin{equation}\label{eq:prob}
p(o)|do|=f(\theta)|d\theta|.
\end{equation}
The density function $f(\theta)$ will depend on details of the theory as well as whatever the resolution of the measure problem may be, but even at this point Eq.~\eqref{eq:prob} shows that a stationary point in $o(\theta)$ will give rise to a spike in $p(o)$ and so, with only mild assumptions as to the form of $f(\theta)$, one can obtain a sharp prediction for the model. Furthermore, if the surface of evaluation is closed, then $o(\theta)$ will be periodic and hence the existence of stationary points is guaranteed! 

At this point some justification is in order as to why it has been decided to work with such a restrictive subclass of multifield models. As will be discussed in more detail in~\S\ref{sec:summary}, the argument above for  the existence of a sharp prediction does not depend on the absence of couplings between the fields, nor on the choice of surface of evaluation. So one should expect that the conclusions reached in this paper by analytic means (achieved by working in this restricted class), could be obtained for a more general class. However one would almost certainly need to resort to numerical techniques to do so, hence it is for the sake of clarity that only this restricted class is considered here.

With regard to relevant work existing in the literature, there seems to be little seeking to directly address the issue of predictions in multifield models. In Refs.~\cite{Frazer:2011tg,Frazer:2011br} a toy model of inflation in the landscape was investigated. By using a method equivalent to taking a flat distribution over initial conditions, distributions for observables were computed. However, the presence of multiple minima meant that the results were clearly sensitive to the choice of distribution of initial conditions yet it was not clear how to asses the impact of this choice on results. A similar approach was taken in Refs.~\cite{Agarwal:2011wm,Dias:2012nf,McAllister:2012am} to study 6-field D-brane inflation but this model suffered from much the same problem. Here, drastically simpler models are considered and the approach taken to studying these models is very different. The most important difference is that here the role of the initial density function is explicit. 

Despite not being directly about inflation, it is interesting to note the recent work of Sumitomo and Tye on the cosmological constant \cite{Sumitomo:2012wa, Sumitomo:2012vx, Sumitomo:2012cf}. The mechanism by which they argue for the smallness of the cosmological constant is closely related to why a spike in the density function of observables is argued to be generic here. 

The rest of this paper is structured as follows. \S\ref{sec:sep} briefly reviews the relevant expressions for observable quantities in sum-separable potentials and discusses the conditions under which the method sketched above is applicable. \S\ref{sec:pfroms} is dedicated to a more thorough explanation of the use of Eq.~\eqref{eq:prob}. \S\ref{sec:quad} is a detailed discussion of the prediction of quadratic inflation; as well as being an interesting model in its own right, the spherical symmetry of the model makes calculating predictions exceptionally straightforward and hence will serve as a transparent demonstration of the method. ~\S\ref{sec:summary} returns to the discussion of why the results should hold for a more general class of models than just the class of sum-separable models and \S\ref{sec:conclusion} concludes this paper.

\section{Sum-Separable Potentials and the Horizon Crossing Approximation}\label{sec:sep}
This section briefly introduces the expressions for observables in canonical models of inflation with sum-separable potentials and then discusses the conditions necessary for the proposed method to be applicable.

\subsection{Expressions for Observables}
The curvature perturbation may be related to perturbations in the fields by realising that on large scales $\zeta$ is equivalent to the perturbation of the number of $e$-foldings from an initial flat hypersurface at $t=t_*$, to a final uniform-density hypersurface at $t=t_{c}$ \cite{Starobinsky:1986fxa,Lyth:1984gv,Sasaki:1995aw,Lyth:2005fi}
\begin{equation}
\zeta(t_c,x) \simeq \delta N(t_c,t_*,x)\equiv N(t_c,t_*,x)-N(t_c,t_*).
\end{equation}
where $N(t_c,t_*) \equiv \int_{*}^{c}H dt$. Expanding $\delta N$ in terms of the initial field perturbations to second order, one obtains
\begin{equation}\label{eq:zeta}
\zeta(t_c,x)=\delta N(t_c,t_*,x)=N,_{\alpha}\delta\phi_\alpha^* +\frac{1}{2}N,_{\alpha\beta}\delta\phi_{\alpha}^*\delta\phi_{\beta}^*,
\end{equation}
where repeated indices should be summed over, and $N,_{\alpha}$, $N,_{\alpha\beta}$ represent first and second derivatives of the number of $e$-folds with respect to the fields $\phi_{\alpha}^*$. 

Observables of interest are related to the correlation functions of $\zeta$ and hence may be expressed in terms of field correlation functions multiplied by derivatives of $N$. In general computing the $N$ derivatives require numerical techniques but a useful exception is the case of sum-separable potentials
\begin{equation}
W(\phi_{1},..,\phi_{N_{\rm f}})=\sum_{i}^{N_{\rm f}}V_{i}(\phi_{i}).
\end{equation}
Models of this kind have the appealing characteristic that observables such as the power spectrum, spectral index, running, non-Gaussianity parameter $f_{\rm NL}$ etc. may be computed analytically. It should be noted however that there is no particularly good reason to believe that these models are well motivated from an effective field theory perspective; one in general should expect couplings between the fields. Although a  more general approach is not given here, in later sections it will be argued that the main result of this paper applies to a broader class of models than just those with sum-separable potentials.

This paper focusses on quantities relating to the two-point statistics of $\zeta$, particularly the spectral index $n_{s}$ and running $\alpha$, however all techniques used throughout this work are just as easily applied to higher order statistics. The reader is referred to Refs.~\cite{Battefeld:2006sz,Vernizzi:2006ve,Dias:2012nf} for derivations of the expressions. For this paper it will suffice to quote the results. Setting $M_{\rm PL}=1$, the power spectrum is given by

\begin{equation}
\label{eq:Pzeta}
{\cal P_{\zeta}}=\frac{W_{*}}{24 \pi^2}\sum_{\alpha=1}^{N_{\rm F}}\frac{u_{\alpha}^2}{\epsilon^{*}_{\alpha}}
\end{equation}
where 
\begin{equation}
u_{\alpha}\equiv\frac{V^{*}_{\alpha}+Z_{\alpha}}{W_{*}},
\end{equation}
and $\epsilon_{\alpha}$ is the first slow-roll parameter
\begin{equation}
\epsilon_{\alpha}\equiv\frac{1}{2}\left(\frac{V_{\alpha}'}{W}\right)^2,
\end{equation}
such that $\epsilon=\sum\epsilon_{\alpha}$. Also, rather importantly for what follows, the term
\begin{equation}\label{eq:Z}
Z_{\alpha}\equiv \frac{1}{\epsilon^{c}}\sum^{N_{\rm F}}_{\beta=1}V_{\beta}^{c}(\epsilon_{\alpha}^{c}-\epsilon^{c}\delta_{\alpha\beta})
\end{equation}
contains all the information about the constant density surface at the time of evaluation and all other terms are evaluated at horizon-crossing.

By differentiating with respect to $\ln k$, the spectral index is found to be 
\begin{equation}\label{eq:n}
n_{s}-1=-2\epsilon_{*}-4\frac{\left(1-\sum_{\alpha=1}^{N_{\rm F}}\frac{\eta^{*}_{\alpha}u_{\alpha}^2}{2\epsilon^{*}_{\alpha}}\right)}{\sum_{\alpha=1}^{N_{\rm F}}\frac{u_{\alpha}^2}{\epsilon^{*}_{\alpha}}},
\end{equation}
where $\eta_{\alpha}$ is the second slow-roll parameter
\begin{equation}
\eta_{\alpha}\equiv\frac{V_{\alpha}''}{W}.
\end{equation}

Differentiating with respect to $\ln k$ once again, the running is found to be \cite{Dias:2012nf}
\begin{equation}\label{eq:run}
\alpha=-8\epsilon^{*2}
+4\sum_{\alpha=1}^{N_{\rm F}} \epsilon^{*}_{\alpha}\eta^{*}_{\alpha} 
-16\frac{\left(1-\sum_{\alpha}\frac{\eta^{*}_{\alpha}u_{\alpha}^2}{2\epsilon^{*}_{\alpha}}\right)^2}{\left(\sum_{\alpha}\frac{u_{\alpha}^2}{\epsilon^{*}_{\alpha}}\right)^2}
-8\frac{\sum_{\alpha}\eta^{*}_{\alpha}u_{\alpha}\left(1-\frac{\eta^{*}_{\alpha}u_{\alpha}}{2\epsilon^{*}_{\alpha}}\right)}{\sum_{\alpha}\frac{u_{\alpha}^2}{\epsilon^{*}_{\alpha}}} 
+4\epsilon^{*}\frac{\sum_{\alpha}\frac{\eta^{*}_{\alpha}u^{2}_{\alpha}}{\epsilon^{*}_{\alpha}}}{\sum_{\alpha}\frac{u_{\alpha}^2}{\epsilon^{*}_{\alpha}}}-2\frac{\sum_{\alpha}\frac{\xi^{*}_{\alpha}u^2_{\alpha}}{\epsilon^{*}_{\alpha}}}{\sum_{\alpha}\frac{u_{\alpha}^2}{\epsilon^{*}_{\alpha}}},
\end{equation}
where $\xi_{\alpha}$ is the third slow-roll parameter
\begin{equation}
\xi_{\alpha} \equiv \frac{V_{\alpha}'V_{\alpha}'''}{W^{2}}.
\end{equation}
All of the above expressions can be shown to reduce to the standard single-field formula by setting $u_{\alpha}=1$.

\subsection{The horizon crossing approximation and the adiabatic limit}
As already discussed, in order to achieve analytic control, the method proposed here for mapping a density function in field space to a density function for observables makes use of the horizon crossing approximation. The above expressions for the power spectrum \eqref{eq:Pzeta}, tilt \eqref{eq:n} and running \eqref{eq:run}, are all composed of slow-roll terms evaluated at horizon crossing ``*" and one other term $u_{\alpha}$, of which the only contribution not evaluated on ``*" is $Z_{\alpha}$. It follows that in order for this approach to be appropriate two requirements must be satisfied:
\begin{enumerate}
\item The slow-roll approximations $\epsilon\ll1$, $\eta\ll1$, $\xi\ll1$ must be valid.
\item Observables must stop evolving within the realm of validity of the model.
\end{enumerate}
The first requirement is necessary for a number of reasons. The method proposed here seeks to map the density function on an $N_{\rm f}-1$ hypersurface in field space to a density function for observables. The very fact that the objective is phrased this way already implicitly assumes that (up to a time shift) there is a unique trajectory passing through each point in field space. A simple way to ensure this is the case is to stipulate that the slow-roll conditions are satisfied. If one does not wish to place any constraints on momenta then the situation is more complicated. In principle an infinite number of both inflationary and non-inflationary trajectories may pass through any given point in field space. To handle such a situation is beyond the scope of this approach but is not a problem in principle. One would just need to consider a suitable slicing of the full phase-space such as constant energy slicing. The reader is referred to Ref.~\cite{Easther:2013bga} for a nice discussion of this matter.

Another reason for the first requirement is that the expressions for observables given above are obtained by assuming slow-roll. In practice it can well be the case that these expressions are valid even when the slow-roll approximations start to break down but this range of validity is clearly model-dependant.

The second requirement refers to the fact that for the model to be predictive, the primordial curvature perturbation must become conserved before the end of inflation. Otherwise a description of reheating is required. This issue applies to all multifield models, not just the sum-seperable models considered here. In order for the curvature perturbation to become conserved, isocurvature modes must have exponentially decayed by the end of inflation such that the model approaches the adiabatic limit \cite{GarciaBellido:1995qq,Elliston:2011dr, Frazer:2011br, Dias:2012nf, Seery:2012vj}. Although it is difficult to say with certainty when isocurvature modes are sufficiently decayed, an intuitive method to test that this requirement has been satisfied is to track the evolution of the width of the bundle of perturbed inflationary trajectories in field space. The reader is referred to Ref.~\cite{Seery:2012vj} for detailed discussion but an expression for the bundle width can be found to be
\begin{equation}
		\Theta(N,N_0)
			= \exp\left\{ -\sum_{\alpha} \int_{\phi_{*}}^{\phi_{c}}(-\eta_{\alpha} +2\epsilon_{\alpha})\frac{V_{\alpha}}{V_{\alpha}'} d\phi_{\alpha}\right\},
		\label{eq:focusing-def}
	\end{equation}
where the expression given here differs from that given in Ref.~\cite{Seery:2012vj} slightly since this has been written for the case of sum-separable potentials and made use of $dN=-\sum_{\alpha}V_{\alpha}/V_{\alpha}' d\phi_{\alpha}$. For an adiabatic limit to be reached, a necessary (but not sufficient if $N_{\rm f}>2$) condition is that $\Theta\rightarrow0$. By inspection of Eq.~\eqref{eq:focusing-def}, a period of  focusing requires 
\begin{equation}
\sum_{\alpha}\eta_{\alpha}>2\epsilon,
\end{equation}
though how much focussing is required is a model dependant statement.

Satisfaction of the second requirement guarantees $Z_{\alpha}$ will be exponentially close to a constant at the end of inflation but does not guarantee that it is negligible. Satisfaction of both the first and the second requirement does however guarantee that to a good approximation one can express observables solely in terms of quantities evaluated at horizon crossing. This is what is referred to as the horizon crossing approximation.\footnote{N.B this is not the same as evaluating the expression for a given observable at horizon crossing!} In practice, it is often the case that $Z_{\alpha}$ becomes negligeable in the approach to the adiabatic limit but even in the situations where this is not the case, provided the first requirement is satisfied, it will be possible to map a given location on the final surface to a unique location on the horizon crossing surface.

\begin{figure}[t]
\centering
\includegraphics[width=10cm]{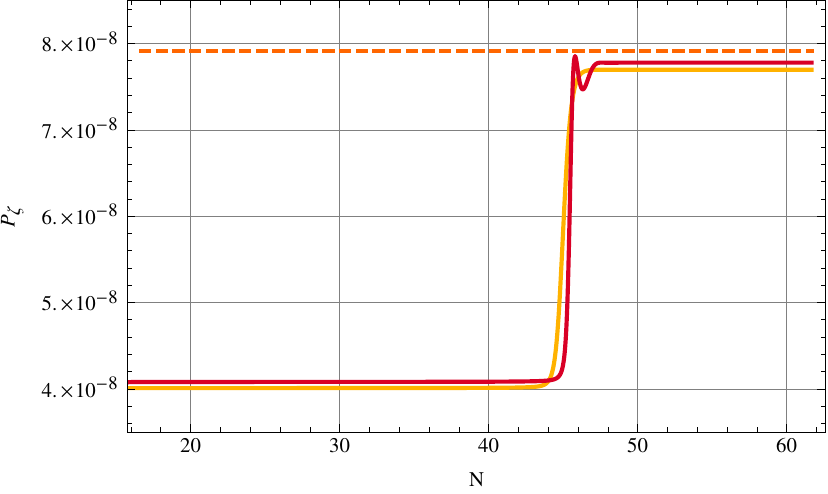}
\caption{Example of the evolution of $P_{\zeta}$ for scales leaving the horizon 55 $e$-folds before the end of inflation in double quadratic inflation with masses $m_{2}/m_{1}=9$ and horizon crossing corresponding to the approximate field-space coordinates $(\phi_{1}^{*}=10.283, \phi_{2}^{*}=10.462)$. The red line shows the non-slow-roll evolution, the gold line shows the evolution using the slow-roll equations and the dashed orange line is the result obtained using the horizon crossing approximation.}
\label{fig:pzetatest}
\end{figure}

For the example of quadratic inflation studied in this paper, $Z_{\alpha}$ is sufficiently small at the end of inflation that it can be dropped from the expressions for observables. In this work, the largest mass ratio explored is $m_{2}/m_{1}=9$ and hence it is in this case that the horizon crossing approximation will be least effective in reproducing the correct result. Fig.~\ref{fig:pzetatest} compares the horizon crossing approximation (dashed orange line) with the full superhorizon evolution computed numerically by use of the transport method. The gold line shows the superhorizon evolution obtained using transport equations derived from the slow-roll equations of motion \cite{Mulryne:2009kh,Mulryne:2010rp}. The red line shows the evolution obtained using the full non-slow-roll transport equations, as described in Ref.~\cite{Seery:2012vj}. Despite the slightly unfortunate name, the horizon crossing approximation does reasonably well in accounting for the non-trivial evolution that occurs due to a sharp turn in the inflationary trajectory. This approximation will fare even better in examples with a less severe mass ratio.

\section{Computing the density function of observables}
\label{sec:pfroms}

The objective is to obtain an expression for $p(o_{1},..,o_{n})$ where $o_{i}$ are the observables of interest such as the spectral index, running etc. If the model consists of $N_{\rm f}$ fields, then the horizon crossing contour is an $(N_{\rm f}-1)$-dimensional hypersurface in field space and hence, under the horizon crossing approximation, each observable may be expressed in terms of $N_{\rm f}-1$ parameters $\theta_{\alpha'}$. Latin indices label the observables, greek indices label coordinates on field space and primed greek indices label coordinates on the horizon crossing hypersurface in field space.  The choice of parameterisation is arbitrary. Describing the horizon crossing hypersurface parametrically such that $\phi_{\alpha}=\phi_{\alpha}(\theta_{1},..,\theta_{N_{\rm f-1}})$, the volume element on the horizon crossing hypersurface can be expressed in terms of the parameters $\theta_{\alpha'}$ in the usual way. Constructing basis vectors
\begin{equation}\label{eq:jacobians}
e^\alpha_{\alpha'}\equiv\frac{\partial \phi_{\alpha}}{\partial \theta_{\alpha'}},
\end{equation} 
assuming flat field space, the induced metric is then
\begin{equation}\label{eq:metrics}
g^{\Sigma_{\rm f}}_{\alpha'\beta'}=\sum^{N_{\rm f}}_{\alpha}e^{\alpha}_{\alpha'}e^{\alpha}_{\beta'}\end{equation}
so remembering that the volume of the corresponding parallelepiped is given by $\sqrt{\det g^{\Sigma_{\rm f}}}$, the volume element $d\theta$ in Eq.~\eqref{eq:prob} can now be written more explicitly as
\begin{equation}
 d\theta\rightarrow\sqrt{\det g^{\Sigma_{\rm f}}}d\theta_{1}\cdots d\theta_{N_{\rm f}-1},
\end{equation}

If $n\leq N_{\rm f}$ then a similar picture holds for the left hand side of Eq.~\eqref{eq:prob}. $o_{i}(\theta_{1},..,\theta_{N_{{\rm f}-1}})$ are the parametric equations of a surface in the space of observables and so one can construct the volume element in the same way. The induced metric in ``observables space" is 
\begin{equation}
g^{o}_{\alpha'\beta'}=\sum^{n}_{i}\frac{\partial o_{i}}{\partial x_{\alpha'}}\frac{\partial o_{i}}{\partial x_{\beta'}}
\end{equation}
and so Eq.~\eqref{eq:prob} becomes

\begin{equation}\label{eq:probjac}
p(o_{1},..,o_{n})=f[\phi_{1}(\theta_{1},..,{{\theta_{N_{\rm f}-1}}}),..,\phi_{N_{\rm f}}(\theta_{1},..,{\theta_{{N_{\rm f}-1}}})]\sqrt{\frac{\det g^{\Sigma_{\rm f}}}{\det g^{\Sigma_{\rm o}}}}.
\end{equation}
 
Although it will always be possible to compute this expression when the horizon crossing approximation is valid and $n\leq N_{\rm f}$, clearly some sacrifices are made to achieve this generality. Ideally what one would like to do is perform the inverse mapping and express $\theta_{\alpha'}(o_{1},...,o_{n})$ such that $p(o_{1},..,o_{n})$ can be written explicitly in terms of the observables. Whether this is practical is a model dependant statement and the difficulty in doing so will also depend on the number of observables under consideration. For instance if $n> N_{\rm f}$, then one either needs to compute the cumulative distribution function or construct enough dummy variables such that $n= N_{\rm f}$. To give an explicit example, consider the case of 1 observable $n=1$ and a 2-dimensional horizon crossing surface $N_{\rm f}-1=2$. One would essentially need to compute
\begin{equation}
p(o)=\frac{d}{do}\iint f\sqrt{det g^{\theta}}d\theta_{1}d\theta_{2},
\end{equation}
which in general one probably would prefer to avoid in favour of numerical techniques.

However Eq.~\eqref{eq:probjac} does have some advantages. As already mentioned, it is simple to compute and still allows one to locate and compare peaks in the density function. In some cases, it is sufficient to construct the full density function and even make a sensible estimate of confidence limits. The main advantage as far as this paper is concerned is that  Eq.~\eqref{eq:probjac} makes manifest the relationship between the geometry of the potential and the initial density function. One potentially valuable application of this is in understanding the sensitivity of the model to the choice of measure. This will be discussed in the context of the example in the following section.

\section{Quadratic Inflation}
\label{sec:quad}

In this section an explicit example is given of the method outlined in \S\ref{sec:pfroms}. Double quadratic inflation is a good example for a number of reasons. Explicitly shown in Ref.~\cite{Seery:2012vj} for an arbitrary number of fields approaching a quadratic minimum, all isocurvature modes are exponentially decaying before the end of inflation and so the horizon crossing approximation is valid. The spherical symmetry of the horizon crossing surface also makes it a particularly simple example.

\begin{figure}[t]
\centering
\includegraphics[width=15cm]{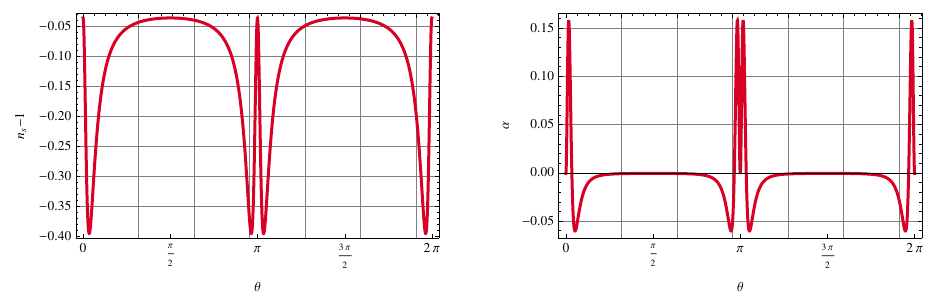}
\caption{Example plots of $n_{s}(\theta)$ and $\alpha(\theta)$ for scales leaving the horizon 55 $e$-folds before the end of inflation in double quadratic inflation with masses $m_{2}/m_{1}=9$.}
\label{fig:naoftheta}
\end{figure}

The potential for double quadratic inflation is \cite{Polarski:1992dq}
\begin{equation}
V_{1}=\frac{1}{2}m_{1}^{2}\phi_{1}^{2}\quad\quad\quad\quad V_{2}=\frac{1}{2}m_{2}^{2}\phi_{2}^{2},
\end{equation}
such that $W=V_{1}+V_{2}$. An expression for the horizon crossing contour may be obtained by writing the number of $e$-folds as
\begin{equation}\label{eq:N}
N(t_{c},t_{*})=-\int_{*}^{c}\sum_{\alpha=1}^{N_{\rm F}}\frac{V_{\alpha}}{V_{\alpha}'}d\phi_{\alpha}
\end{equation}
and so
\begin{equation}
N=\frac{1}{4}(\phi_{1*}^{2}+\phi_{2*}^{2})-\frac{1}{4}(\phi_{1c}^{2}+\phi_{2c}^{2}).
\end{equation}
A well known \cite{Vernizzi:2006ve,Polarski:1992dq} and rather helpful choice of parameterisation is to move into polar coordinates. By neglecting the contribution from the ``c" surface one can write
\begin{equation}\label{eq:polar}
\phi_{1}=2\sqrt{N}\cos\theta\quad\quad\quad \phi_{2}=2\sqrt{N}\sin\theta,
\end{equation}
where here and what follows, unless explicitly written otherwise, fields only lie on the surface $\Sigma_{\rm f}$ and so the ``*" label has been dropped. For quadratic inflation,  the observables \eqref{eq:Pzeta}, \eqref{eq:n} and \eqref{eq:run} become
\begin{eqnarray}
{\cal P_{\zeta}}(\theta,N)&=& \frac{H^{2}}{4\pi^{2}}N, \\
n_{s}(\theta,N)-1&=& -2\epsilon-\frac{1}{N},\\
\alpha(\theta,N)&=& -8\epsilon^2-\frac{2}{N^2}+4\left(\epsilon_{1}\eta_{1}+\epsilon_{2}\eta_{2}\right).
\end{eqnarray}
The amplitude of the power spectrum is in a sense less interesting than the other observables since it may always be adjusted by a pre-factor on the potential which does not affect the inflationary dynamics. Other observables one might want to consider are the non-Gaussianity parameter $f_{\rm NL}$ and the tensor-to-scalar ratio $r$. For the case of double quadratic inflation, it was shown in Ref.~\cite{Vernizzi:2006ve} that  $-\frac{6}{5}f_{\rm NL}={1}/{N}$. Since $\Sigma_{\rm f}$ is defined as being at a fixed number of $e$-folds before the end of inflation, $f_{\rm NL}$ is single valued over $\Sigma_{f}$ which is why it has not been part of the discussion until now (the same is true for the tensor-to-scalar ratio). Hence, for this example, the objective is to calculate $p(n_{s},\alpha)$. Fig.~\ref{fig:naoftheta} shows plots of the spectral index and running evaluated 55 $e$-folds before the end of inflation. As mentioned previously, an important characteristic to bear in mind for what follows is that these observables are periodic in $\theta$. This is simply a consequence of the fact that $\theta$ represents coordinates on a closed surface. It is also worth noting that the stationary points at $\theta = \pi i/2$, where $i$ is an integer, occur at the same values for $n_{s}$ and $\alpha$.  

\subsection{Regarding the field space density function $f_{\Sigma_{\rm f}}$}
It is necessary to specify the density function $f_{\Sigma_{\rm f}}$. As already mentioned, this is dependant on the details of the model as well as the choice of measure. For the purposes of this paper the distribution is chosen to be flat
\begin{equation}
f_{\Sigma_{\rm f}}=c=\frac{1}{\int_{\Sigma_{\rm f}}d\Sigma_{\rm f}}.
\end{equation}
Following Ref.~\cite{de2007philosophical}, this choice is a statement of ignorance. We simply adopt the distribution requiring the least additional assumptions. No further justification will be given at this stage, however in \S\ref{sec:flatdist} the implications of this choice will be discussed.

The physics in each quadrant of field space is the same, so without loss of generality consider just the first quadrant where both fields are positive. The density function over the contour is then
\begin{equation}
f_{\Sigma_{\rm f}}=\frac{1}{\pi\sqrt{N}}.
\end{equation}

\subsection{The joint density function $p(n_{s},\alpha)$}
As mentioned, it is convenient to parameterise in terms of polar coordinates $\{N,\theta\}$. Pythagorus and Eq.~\eqref{eq:polar} give $d\Sigma_{\rm f}=2\sqrt{N}d\theta$ and so the right hand side of Eq.~\eqref{eq:prob} becomes
\begin{equation}\label{eq:dtheta1}
f_{\Sigma_{\rm f}}d\Sigma_{\rm f}=\frac{2}{\pi}d\theta.
\end{equation}
Since for this model the observables of interest are $o=\{n_{s},\alpha\}$, using Eq.~\eqref{eq:jacobians} and Eq.~\eqref{eq:metrics} the element $do$ may be rewritten as
\begin{equation}\label{eq:do1}
do=\sqrt{\left(\frac{d n_{s}}{d\theta}\right)^2+\left(\frac{d \alpha}{d\theta}\right)^2}d\theta
\end{equation}
and hence combing Eq.~\eqref{eq:prob} with Eq.~\eqref{eq:dtheta1} and  Eq.~\eqref{eq:do1},
\begin{equation}
p(n_{s},\alpha)=\frac{2}{\pi}\frac{1}{\sqrt{\left(\frac{d n_{s}}{d\theta}\right)^2+\left(\frac{d \alpha}{d\theta}\right)^2}}.
\end{equation}

\begin{figure}[t]
\centering
\includegraphics[width=15cm]{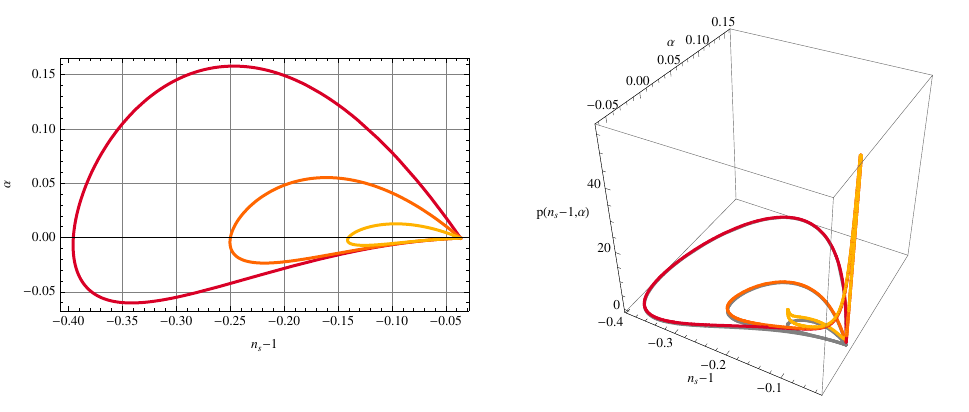}
\caption{Example plots for double quadratic potential with masses $m_{2}/m_{1}=9$ (red), $m_{2}/m_{1}=7$ (orange), $m_{2}/m_{1}=5$ (gold). The plot on the left shows the set of possible values of $\{n_{s},\alpha\}$. The right hand plot additionally includes $p(n_{s},\alpha)$. The density function for all examples shown peak at $n_{s}-1=-\frac{1}{30}$ and $\alpha=-\frac{1}{1800}$ and are divergent and so the full plot range is not shown. The grey lines are the projections onto the $p(n_{s},\alpha)=0$ plane.}
\label{pofna}
\end{figure}

The right hand plot of Fig.~\ref{pofna} shows $p(n_{s},\alpha)$ for the case of double quadratic inflation with masses $m_{2}/m_{1}=9$, $m_{2}/m_{1}=7$, $m_{2}/m_{1}=5$ shown in red, orange and gold respectively. The left plot shows the contours of possible values in the $n_{s}-\alpha$ plane. There is a strong peak in the density function which corresponds $\theta=\pi i/2$, where $i$ is a member of the integers. Comparing with Fig.~\ref{fig:naoftheta}, this is exactly what should be expected. Fig.~\ref{fig:naoftheta} shows that for a large range of $\theta$, the tilt and running are slowly varying. Since a large proportion of the contour maps to a relatively small proportion of the possible range of values of the tilt and running, this will give rise to a very sharp spike in the density function.  

It is interesting to note that the peak occurs at the same point for each mass ratio. i.e the prediction of the model is not as sensitive to the mass hierarchy as the space of possible values for observables might suggest.

\section{Discussion}\label{sec:summary}
In computing the density function for observables in the previous section, two important assumptions were made. One was that the density function on field space was taken to be a flat distribution over the horizon crossing contour. The second was that the potential was sum-separable. In this section the significance of these assumptions is assessed.
\begin{figure}[t]
\centering
\includegraphics[width=15cm]{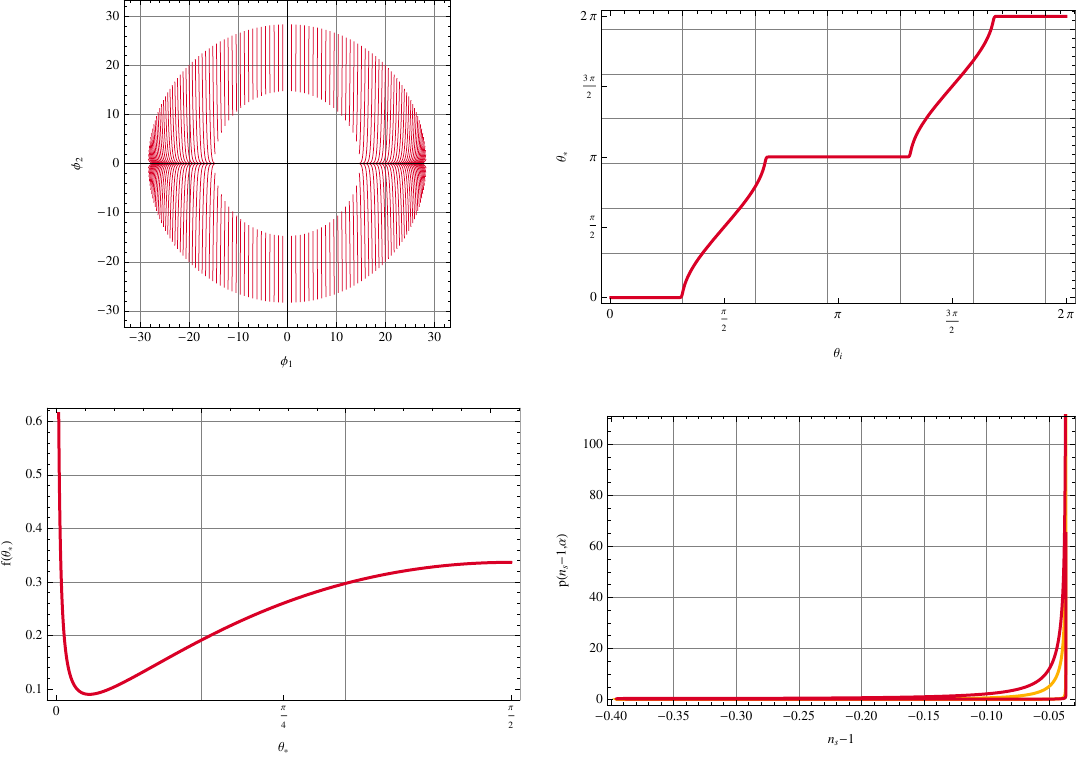}
\caption{Example plots showing the effect on $p(n_{s}-1,\alpha)$ if a flat distribution is taken at 200 $e$-folds before the end of inflation instead of at horizon crossing, for the case of double quadratic inflation with mass ratio $m_{2}/m_{1}=9$. The top left plot shows the evolution of example trajectories from 200 $e$-folds before the end of inflation up until horizon crossing. The plot shows a significant proportion of the trajectories converge at $\theta=0$ and $\theta=\pi$. The top right plot shows the relation of $\theta_{i}$, the parameterisation of the initial contour, to $\theta_{*}$, the parameterisation of the horizon crossing contour. The plot shows once again that trajectories over a significant range of $\theta_{i}$ converge onto $\theta_{*}=0$ and $\theta_{*}=\pi$. The bottom left hand plot shows the resulting density function over one quadrant of the horizon crossing surface with a strong peak at $\theta=0$. Finally, the bottom right plot shows the result on $p(n_{s}-1,\alpha)$ in gold compared to $p(n_{s}-1,\alpha)$ for a flat distribution at horizon crossing shown in red. The red plot is the same as that shown in the right hand plot of Fig.~\ref{pofna} but projected onto the $p(n_{s}-1,\alpha)-n_{s}-1$ plane. The effect of the evolution is to sharpen the peak in $p(n_{s}-1,\alpha)-n_{s}-1$.}
\label{thetadiscuss}
\end{figure}

\subsection{The choice of a flat density function at horizon crossing}\label{sec:flatdist}
Following the procedure in Ref.~\cite{Polarski:1992dq}, an expression for the evolution of $\theta$ from some initial surface $\Sigma_{i}$ to the horizon crossing surface $\Sigma_{*}$ can be obtained by substituting Eq.~\eqref{eq:polar} into the slow-roll equations of motion for the fields and integrating 
\begin{equation}
\frac{N_{*}}{N_{i}}=\left(\frac{\sin\theta_{*}}{\sin\theta_{i}}\right)^{\frac{2m_{1}^2}{m_{2}^{2}-m_{1}^{2}}}\left(\frac{\cos\theta_{*}}{\cos\theta_{i}}\right)^{\frac{2m_{2}^2}{m_{2}^{2}-m_{1}^{2}}}.
\end{equation}
As shown by the example plots in Fig.~\ref{thetadiscuss}, there is a dynamical attractor causing the field trajectories to converge on $\theta=0$ or $\theta=\pi$. This means that a flat distribution chosen on some $N>N_{*}$ contour will evolve to give rise to a density function at horizon crossing with peaks at $\theta=0$ and $\theta=\pi$. As shown in Fig.~\ref{thetadiscuss}, the result of this is to sharpen the peak in the density function of observables. In this sense choosing a flat distribution at horizon crossing can be considered to result in a lower bound on the strength of the peak sourced by dynamical effects.

As mentioned previously, one advantage of computing the density function as given by Eq.~\eqref{eq:probjac}, is that it makes clear the role of the density function on field space and so in turn, the choice of measure. Since field space dynamics act to strengthen the peak in $p(o)$, in order for the prediction of the model to be changed significantly, one would require a choice of measure that acts to counter this dynamical effect. 

While beyond the scope of this paper to study such a model, Hybrid inflation \cite{Kofman:1986wm,Linde:1993cn,Copeland:1994vg} represents an interresting example of a model which may lend a rather different perspective to the above the discussion. In this model, it has been shown that the set of initial conditions giving rise to sufficient inflation has a fractal structure \cite{Easther:1997hm,Ramos:2001zw,Clesse:2009ur,Lazarides:1996rk,Lazarides:1997vv,Tetradis:1997kp,Mendes:2000sq,Clesse:2008pf,Easther:2013bga}. This characteristic stems from the fact the inflationary region constitutes a small fraction of the field space but is also an attractor, and hence a large proportion of the non-inflationary initial conditions will eventually seek out the inflationary region. For this model, well behaved constant $e$-fold surfaces still exist but one can imagine that different priors on the phase space might correspond to a highly non-trivial density function on a given constant $e$-fold hypersurface. While in practice this is of no issue due to the fact that the model is already highly predictive (the full phase space maps to a small volume in the space of observables), it does suggest that more complicated models might be easier to study using a different slicing such as, for instance, constant energy slicing\footnote{I would like to thank Layne Price and Richard Easther for helpful discussions on this point.} (see Ref.~\cite{Easther:2013bga} for a detailed discussion of the advantages of this choice).

\subsection{The use of sum-separable potentials}

Central to the approach taken in this paper is the use of sum-separable potentials. It is therefore important to understand to what extent the results are special to this class of models. The reason the analysis performed here was restricted to this class was that it enabled the use of the horizon crossing approximation and hence provided a simple analytic method of mapping a density function on field space to a density function for observables. The most striking result is that the functional form of  Eq.~\eqref{eq:probjac} seems to imply that the density function of observables will be strongly peaked and this is indeed the case for the example of double quadratic inflation that was explored. If a strong peak turns out to be a generic feature of multifield models this is advantageous on two fronts. Most importantly it means that a given multifield model may be significantly more predictive than considering just the range of possible values of observables allowed within the model and hence one can be much more optimistic about the prospect of constraining such models. Secondly, if sharp peaks in the density function exist, then knowing the location of these peaks can be very advantageous to numerical approaches to computing the density function. 

Although the use of the horizon crossing approximation provides a simple map from field space to observables, it is argued here that it is not the cause of the peak in $p(o)$. It may not be very efficient but clearly there already exists a method of computing the density function of observables for a much broader class inflationary models than the method given here. For example, consider the class of all inflationary models where the slow-roll approximation is valid at horizon crossing and also reach an adiabatic limit before the end of inflation.\footnote{The method proposed here is straightforward to adapt to other models of inflation but some details may vary and so in pursuit of being a little more concrete, discussion has been restricted to this class.} For this class one could perform the following procedure:

\begin{enumerate}
\item Perform a Monte Carlo search for the horizon crossing surface $\Sigma_{*}$.
\item Draw initial field space positions from a given density function $f_{\Sigma_{*}}$.
\item For each set of initial field space positions, starting with momenta given by the slow-roll equations, evolve each trajectory until the end of inflation.
\item For each inflationary trajectory, using one of the methods given in Refs.~\cite{Starobinsky:1986fxa,Lyth:1984gv,Sasaki:1995aw,
	Salopek:1990jq,Sasaki:1998ug,Wands:2000dp,Lyth:2005fi,Rigopoulos:2004gr,Rigopoulos:2005xx,Yokoyama:2007uu,Yokoyama:2007dw,Yokoyama:2008by,Mulryne:2009kh,Mulryne:2010rp,Amendola:2001ni,GrootNibbelink:2001qt,Lalak:2007vi,Peterson:2010np,
	Peterson:2010mv,Peterson:2011yt,Achucarro:2010da,Avgoustidis:2011em,Lehners:2009ja,Ringeval:2007am,Martin:2006rs,Huston:2009ac,Huston:2011vt}, compute relevant observables.
\end{enumerate}

In doing this, one would numerically obtain a map from the $N_{\rm f}-1$ dimensional surface on which inflation ends to observables which may, as before, be parameterised by $N_{\rm f}-1$ variables $\{\theta_{1},..,\theta_{N_{{\rm f}-1}}\}$. Once this map is obtained, Eq.~\eqref{eq:probjac} is once again applicable. 

If the surface in field space is closed there are only two options for the mapping. Either the observable is independent of $\{\theta_{1},..,\theta_{N_{{\rm f}-1}}\}$, or it is periodic in $\{\theta_{1},..,\theta_{N_{{\rm f}-1}}\}$. If the observable is independent of $\{\theta_{1},..,\theta_{N_{{\rm f}-1}}\}$ then (as was the case here for $f_{\rm NL}$ and  $r$) that observable is single valued. If the observable is periodic in $\{\theta_{1},..,\theta_{N_{{\rm f}-1}}\}$, then stationary points are guaranteed.

\begin{figure}[t]
\centering
\includegraphics[width=10cm]{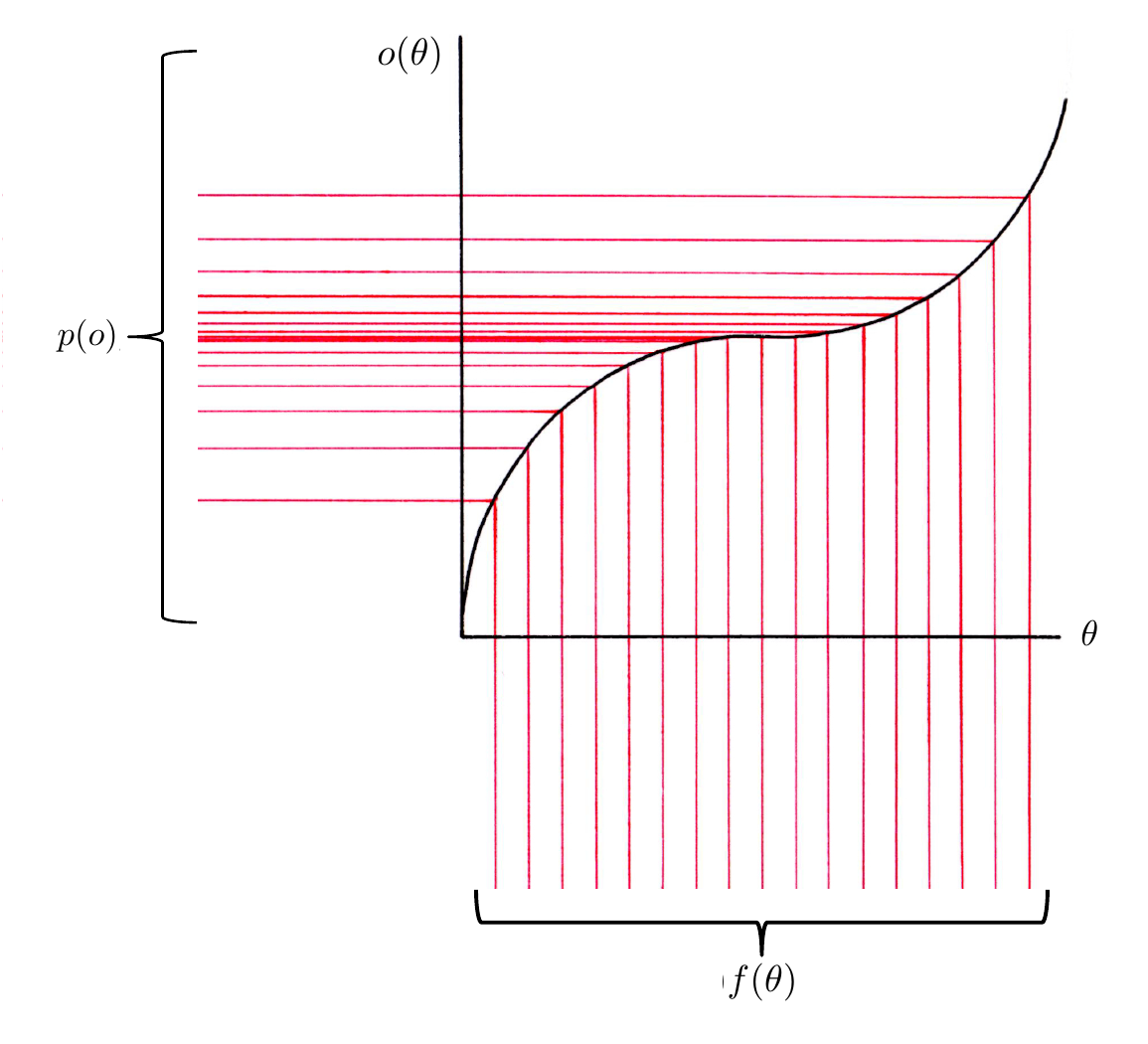}
\caption{Sketch of why a stationary point in the functional form of $o(\theta)$ will give rise to a divergence in the density function $p(o)$. The spacing of the red lines represents the density function $f(\theta)$ and how it is distorted under a change of variable to give $p(o)$.}
\label{density}
\end{figure}

If the surface in field space is not closed then it is not possible to say anything so concrete. Nevertheless, the conditions for a peak in the density function to occur are clear. As illustrated in Fig.~\ref{density}, a peak in $p(o)$ occurs when $do/d\theta$ is small. If the surface in field space is not closed, then this requirement can not be guaranteed but it is certainly still permitted.

\section{Conclusion}\label{sec:conclusion}
If multifield models of inflation are to be confronted with data, it is essential that we understand what the prediction of a given model actually is. As has been discussed, there is essentially two parts to this problem:
\newpage
\begin{enumerate}
\item Computing the density function on field space.
\item Mapping the density function on field space to the density function of observables.
\end{enumerate}

Progress with the first part is stymied by the measure problem. This issue has received considerable attention recently and a number of authors have proposed various possible solutions. At present however, the problem remains open. The second part has received comparatively little attention but it is hoped that this paper represents steps towards rectifying this. 

As an attempt at making progress with the second issue, here a solution for the case of sum-separable potentials was proposed. Although it is useful to have analytic examples, clearly this is not sufficient and a more general approach is needed if we are to compare multifield models with data from Planck or any other future survey. As discussed, in principle a numerical approach is already available. However it is inefficient, meaning that a thorough analysis of models with a large number of active fields, or models with complicated Lagrangians may still be computationally unaccessible.

The main point this work is trying to make is that multifield models can have a strongly peaked density function for observables. This means that some models may be relatively insensitive to the density function on field space and hence for such models it should be possible to make robust predictions with the knowledge we already have available to us.

\acknowledgments
I would like to thank Christian Byrnes, Mafalda Dias, Joseph Elliston, David Mulryne, Andrew Liddle and David Seery for numerous exceedingly valuable discussions throughout the course of this work and also for useful comments on early drafts of this paper. I would also like to thank the referee and Layne Price for exceedingly helpful comments on a later draft. Finally, I would like to thank Ground Coffee House in Kemptown, Brighton, where a great deal of this work was done. I am supported by the Science and Technology Facilities Council [grant number ST/I506029/1].

\bibliography{references}

\end{document}